\documentclass[letterpaper,9pt, onecolumn, twoside, nosectionnumbsers]{classes/LTJournalArticle}

\def\doctitle{Saturable absorption in defect-rich diamond nanophotonics}

\def\authorOne{Christopher Coutts}
\def\authorTwo{Nicholas J. Sorensen}
\def\authorThree{Elham Zohari}
\def\authorFour{Sean McNaney}
\def\authorFive{Sigurd Fl{\aa}gan}
\def\authorSix{Paul E. Barclay}
\def\addressOne{
    Institute for Quantum Science and Technology, University of Calgary, Calgary, AB, T2N 1N4, Canada
    }
\def\addressTwo{Department of Physics, University of Alberta, Edmonton, Alberta, T6G 2R3, Canada}
\def\addressThree{National Research Council of Canada, Quantum and Nanotechnology Research Centre, Edmonton, Alberta, T6G 2M9, Canada}
\def\addressFour{The authors contributed equally to this work}

\def\emailContact{christopher.coutts@ucalgary.ca} 

\def\abstractText{
Diamond is a leading quantum photonics platform due to its ability to host qubits based on crystal defects such as nitrogen-vacancy centres. Fabricating nanophotonic devices from defect-rich diamond, which underpins many quantum sensing technologies, promises enhanced performance and integrability of diamond quantum sensors.
Here we demonstrate microdisk cavities fabricated from defect-rich diamond that support optical modes with high quality factor (${Q}\sim7\times10^4$ at $1042\,$nm) and show that they exhibit saturable absorption.  
Power-dependent spectroscopy measurements spanning 979\,nm to 1604\,nm are used to observe intensity-dependent cavity loss and extract wavelength-dependent absorption coefficients and saturation intensities.
At 1047\,nm, we observe saturation and measure a saturation intensity of $2.1\,(8)\,$\,MW/cm$^2$ and an absorption coefficient of $0.53\,(2)$\,cm$^{-1}$. These results provide insight into defect-mediated optical loss in diamond nanophotonics and suggest strategies to harness defect-induced nonlinearities in future diamond photonic devices.
}

\usepackage{graphicx}%
\usepackage{multirow}%
\usepackage{amsmath}%
\usepackage[dvipsnames]{xcolor}%
\usepackage{listings}%
\usepackage{lineno}
\usepackage{ulem}
\usepackage{soul}
\usepackage{upgreek}
\usepackage{braket}

\usepackage{orcidlink,breqn,enumitem,caption,subcaption,ulem,titlesec,lipsum,mathtools,gensymb,float,setspace,siunitx,multicol}

\usepackage[nameinlink]{cleveref}
\usepackage{dirtytalk}
\usepackage{upgreek}
\usepackage{dblfloatfix}

\Crefname{equation}{Eq.}{Eqs.}
\Crefname{figure}{Fig.}{Figs.}
\Crefname{table}{Tab.}{Tabs.}

\begin{document}
\title{\doctitle}

\author{\normalsize
\authorOne,\textsuperscript{1,$\dagger$,*}\orcidlink{0009-0007-7119-9896} 
\authorTwo,\textsuperscript{1,$\dagger$}\orcidlink{0000-0002-4666-5791} 
\authorThree,\textsuperscript{1,2,3} \orcidlink{0000-0001-8344-4157} 
\authorFour,\textsuperscript{1}\\   \normalsize
\authorFive,\textsuperscript{1}\orcidlink{0000-0003-0272-7601} and 
\authorSix\textsuperscript{1}\orcidlink{0000-0002-9659-5883}}
\date{\small\textit{\textsuperscript{1}\addressOne  \\ 
\textsuperscript{2}\addressTwo \\ 
\textsuperscript{3}\addressThree \\ 
\textsuperscript{$\dagger$}\addressFour \\ \textsuperscript{*}\emailContact}}

\twocolumn[
\maketitle
\begin{abstract}
    \abstractText
\end{abstract}
\vspace{10pt}
]

\section{Introduction}

\begin{figure}[t!]
    \centering
    \includegraphics[]{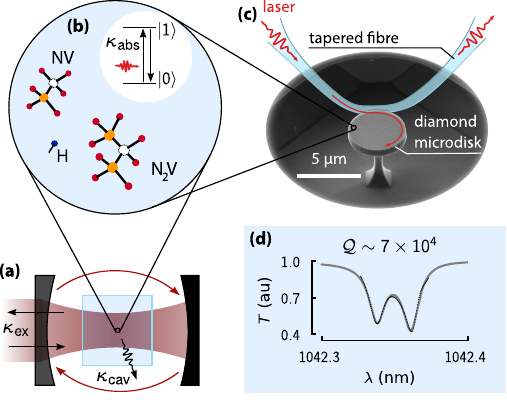}
    \caption{\textbf{Loss in a diamond cavity.} \textbf{(a)} A diamond crystal inside a Fabry-Perot cavity. 
    Photons are side-coupled into the cavity at rate $\kappa_{\text{ex}}$, whose linewidth is determined by the total optical cavity loss, $\kappa = \kappa_{\text{ex}} + \kappa_{\text{cav}}$. 
    Various mechanisms contribute to the cavity loss rate $\kappa_{\text{cav}}$, and in highly doped samples absorption loss from defects is significant. 
    \textbf{(b)} The diamond crystal hosts a variety of different point defects, several of which cause absorption loss ($\kappa_{\text{abs}}$) including hydrogen-based defects (H), and nitrogen-based defects like the NV centre and the $\text{N}_2\text{V}$ centre.
    \textbf{(c)} A scanning electron micrograph reveals the diamond microdisk used to study the absorption dynamics. 
    A fibre-taper waveguide is used to couple light into the microdisk, and changes in the transmission spectrum are used to characterize saturable absorption by the whispering-gallery mode \textbf{(d)}. 
    Fitting the transmission spectrum of the mode at 1042.35\,nm, we find quality factors of ${Q}_{\text{s}}=71.6\,(6)\times10^3$ and ${Q}_{\text{a}}=76.0\,(5)\times10^3$ for the two dips corresponding to the symmetric and antisymmetric standing wave modes of the WGM, respectively.
    }
    \label{fig:fig1_cavityLoss}
\end{figure}

Diamond holds tremendous promise for quantum photonic technologies due to its wide electronic bandgap, its exceptional thermal properties, and its ability to host optically addressable defects that function as spin qubits~\cite{Shandilya2022}.
In addition, diamond cavities can strongly confine light and operate at extreme optical intensities without sustaining optical damage~\cite{Shandilya2024,Itoi2025arXiv}.
Such operating regimes significantly enhance light–matter interactions and, together with advances in fabrication techniques~\cite{Khanaliloo:2015,Castelletto2017,Mitchell:2019,Kuruma2025,Riedel2026}, have enabled diamond nanophotonic cavities for applications including nonlinear optics~\cite{Hausmann2014,Latawiec2015,Itoi2025arXiv,Flagan:2025}, cavity-enhanced spin–photon interactions~\cite{Janitz:2020,Bhaskar2020,Beukers2024}, and optomechanics~\cite{Burek2016,Mitchell:2019,Shandilya:2021}.
Realizing cavities fabricated from material engineered to host dense ensembles of defects is of growing interest for quantum technologies~\cite{Jensen2014}; however, the impact of such defects on the optical properties of nanophotonic devices has not been studied.

As illustrated in \Cref{fig:fig1_cavityLoss}, optical cavities resonantly circulate light and enhance light-matter interactions~\cite{Flagan2022Dres}.
The performance of optical cavities in many applications can be characterized by the ratio of cavity quality factor to effective mode volume, ${Q}/V$, and is inversely proportional to the energy loss rate of the cavity, ${Q}\propto 1/\kappa$ (see Appendix \ref{sec:MicrodiskCMT}). 
Cavity loss arises from several mechanisms that contribute to distinct loss rates: the intrinsic loss rate $\kappa_{\text{i}}$, the external loss rate $\kappa_{\text{ex}}$, the parasitic loss rate $\kappa_{\text{p}}$, and the absorption loss rate $\kappa_{\text{abs}}$. 
Coupling to radiation modes due to fundamental leakage, as well as scattering from fabrication-induced imperfections and surface roughness, contributes to $\kappa_{\text{i}}$. 
Coupling to the output and leaky modes of a waveguide used to input and collect light from the cavity is captured by $\kappa_{\text{ex}}$ and $\kappa_{\text{p}}$, respectively, while optical absorption by defects contributes to $\kappa_{\text{abs}}$.

At low optical intensities ($I$), these loss rates are typically assumed to be power-independent. However, absorption can exhibit nonlinear behaviour at high $I$, leading to an intensity-dependent internal loss rate for the cavity, $\kappa_{\text{cav}}$:
\begin{align}
    \kappa_{\text{cav}}(I) = \kappa_{\text{i}}+ \kappa_{\text{p}}+ \kappa_{\text{abs}}(I)\,.
    \label{eq:nonlinearCavityLoss}
\end{align}

In previous studies of diamond nanophotonic devices, defect-related absorption has not been observed, in part because devices are typically fabricated from high~\cite{Mitchell:2019} or ultrahigh~\cite{Burek2016} purity material and studied at visible or telecom wavelengths.  
However, diamond quantum sensing applications often benefit from samples with high defect density~\cite{Sorensen:2025, ElementSix2021}. 
Of the many optically active defects in diamond~\cite{Aharonovich2011}, including nitrogen-~\cite{Ashfold2020}, hydrogen-~\cite{Czelej2018}, and group-IV-based~\cite{Rose2018,Janitz:2020,Bradac2019} aggregates (see \Cref{fig:fig1_cavityLoss}\,(b)), spin-defects like the nitrogen-vacancy (NV) and the silicon-vacancy have been extensively studied for quantum networking~\cite{Togan:2010,Janitz:2020,Bhaskar2020, Knaut2024,Stolk2024} and sensing~\cite{Rondin2014,Degen2017} applications. 
Furthermore, samples engineered to support high spin-defect densities often host additional undesired defects like the nitrogen-vacancy-nitrogen ($\text{N}_2\text{V}$)\,\cite{Wong:2022} and nitrogen-vacancy-hydrogen (NVH)~\cite{Glover:2003}. 
Because high-${Q}/V$ cavities create large intracavity intensities for modest input powers and exhibit an optical response that is affected by small changes in $\kappa_{\text{cav}}$, they can reveal nonlinear optical processes related to these defects that are not easily observable in bulk samples.

In this work, we realize microdisk cavities fabricated from dense-NV ``quantum-grade'' diamond ([NV]~$\sim$~4.5\,ppm~\cite{ElementSix2021}), an example of which is shown in \Cref{fig:fig1_cavityLoss}~(c).
These devices support optical whispering-gallery modes (WGMs) with effective mode volumes below $8\times(\lambda/n)^3$ and quality factors exceeding $7\times10^{4}$, where $n$ is the refractive index of diamond and $\lambda$ is the vacuum wavelength.  
We observe saturable absorption in these devices for input powers smaller than 10\,mW and isolate its contribution to total cavity loss by measuring the intensity dependence of $\kappa_{\text{abs}}(I)$.
To characterize the nature of the saturable absorption, we perform power-dependent measurements on WGMs between wavelengths of $940$\,nm and $1640$\,nm and observe saturable absorption for modes at wavelengths between 979\,nm and 1267\,nm.
We attribute this previously unobserved effect to the presence of defects in the diamond crystal and discuss its impact on device performance, as well as potential opportunities for dynamic nonlinear photonics~\cite{Dangoisse1990,Paschotta2013} and sensing technologies~\cite{Qu2022}.

\begin{figure*}[t!]
    \centering
    \includegraphics[width = \textwidth]{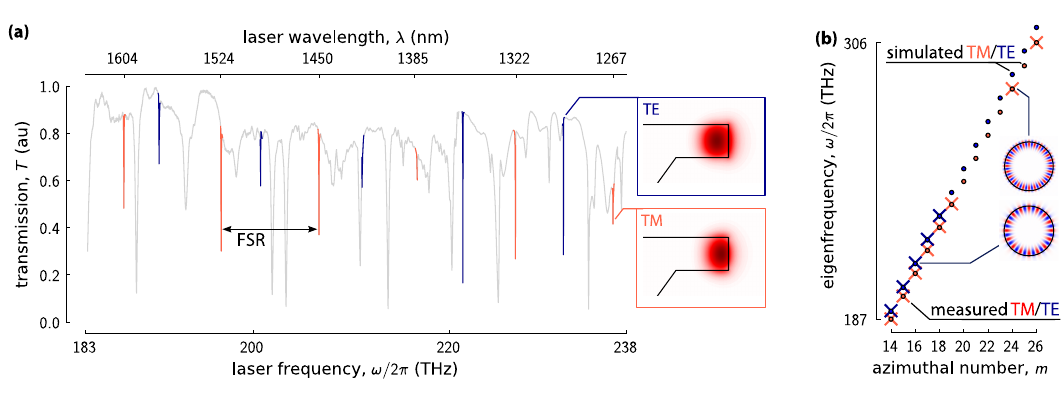}
    \caption{
    \textbf{Fundamental WGMs in a diamond microdisk.}
    \textbf{(a)} A wideband transmission spectrum reveals an array of modes, and we highlight the fundamental TM (TE) modes in red (blue). 
    Eigenmodes of the resonator are simulated using a finite element solver, and insets show the simulated TM and TE mode field distributions (azimuthal mode number $m=18$) inside the microdisk.
    \textbf{(b)} The simulated eigenfrequencies, marked using dot scatter points, align with the measured ones, marked using the `x' scatter points. 
    The azimuthal field distribution of two of the modes are shown as insets to (b).
    Note that the two highest energy TM eigenmode transmission spectra are not shown in (a).}
    \label{fig:fig2_WGModes}
\end{figure*}

\section{Broadband characterization of dense-NV diamond microdisk resonances}

To study optical loss in the dense-NV diamond microdisks, we measure their WGM spectrum using a dimpled fibre-taper waveguide  \cite{Michael2007} positioned adjacent to the cavity, as illustrated in \Cref{fig:fig1_cavityLoss}\,(c) and described in Refs~\cite{Mitchell:2019,Masuda:2024}.
Laser light from continuous-wave tunable lasers (Santec TSL-570 and Newport Velocity TLB-6719 and TLB-6721) spanning $940-1640$\,nm was evanescently coupled into the microdisks from the fibre-taper, whose transmission as a function of wavelength was monitored using a photodetector.
Optical power input to the fibre-taper was varied using an attenuator positioned at the amplifier output while a second attenuator placed immediately prior to the detector ensures that the measured transmitted signal remains within the optimal operating range of the detector.
A typical mode spectrum near 1042 nm---the wavelength often used for absorption based NV magnetometry \cite{Acosta2010, Younesi:2025}---is shown in \Cref{fig:fig1_cavityLoss}~(d), demonstrating $Q \sim 7 \times 10^4$ and a doublet structure resulting from surface-roughness-induced mode coupling \cite{Mitchell:2019}. A transmission spectrum spanning the scan range from $1260-1640$\,nm is shown in \Cref{fig:fig2_WGModes}~(a) and reveals a series of WGMs, all of which are under-coupled. Additional details of the experimental setup are provided in Appendix \ref{sec:ExpSetup}.

To distinguish between fundamental and higher-order WGMs, we simulate the microdisk eigenmodes using the COMSOL Multiphysics finite element solver (FES).
Although the cavity was designed with a 4.2\,$\upmu$m diameter, small fabrication-induced variations inevitably introduce dimensional deviations. In particular, the microdisk thickness obtained from the quasi-isotropic undercut fabrication process varies depending on etching conditions.
To determine the physical dimensions of the fabricated cavity, we simulate microdisks over a range of diameters and thicknesses and identify the geometry whose free spectral range (FSR) and eigenfrequencies best match those identified using the measured transmission spectra~\cite{Behjat2023,Masuda:2024}.
We find that the simulated and measured fundamental eigenfrequencies exhibit optimal agreement for a diamond microdisk with a 4.15\,$\upmu$m diameter and 800\,nm thickness. These dimensions also agree with those extracted from a scanning electron micrograph, within uncertainty. 
The experimentally identified TM (TE) modes are highlighted in red (blue) in \Cref{fig:fig2_WGModes}~(a), and representative cross-sections of the simulated electric field amplitudes for one TM and one TE mode are shown in the inset.
The measured fundamental modes exhibit an FSR of approximately 10\,THz. 
The close agreement between measured and simulated eigenfrequencies, shown in \Cref{fig:fig2_WGModes}~(b), demonstrates the accuracy of the simulated cavity geometry.
Beyond enabling reliable mode identification, the FES simulations provide the effective mode volume and group index for each WGM, both of which are required to calculate the intracavity optical intensity from experimentally measured parameters. Because these quantities are derived from the simulated cavity geometry, their accuracy depends on the fidelity of the geometric model. The strong agreement between measured and simulated eigenfrequencies indicates that the simulation-derived parameters accurately represent the experimental microdisk. Consequently, uncertainties associated with derived quantities such as the effective mode volume are expected to be significantly smaller than uncertainties associated with other experimental parameters, including the absolute laser power and fibre-taper coupling efficiency. Additional details regarding the eigenmode simulations are provided in Appendix \ref{sec:EigenmodeSim}.

\section{Observation of saturable absorption in dense-NV diamond}
\begin{figure}[t!]
    \centering
    \includegraphics[]{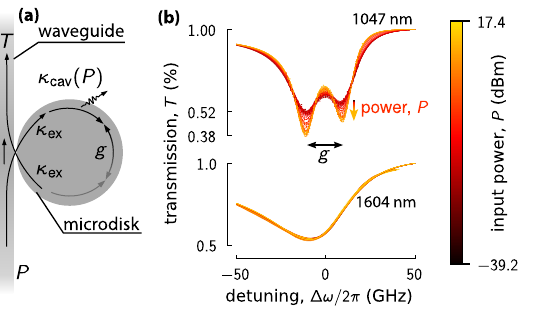}
    \caption{\textbf{Power-dependent laser transmission scans in a microdisk.} 
    \textbf{(a)} Photons coupled into and out of the microdisk at rate $\kappa_{\text{ex}}$, and the internal cavity loss is power-dependent, $\kappa_{\text{cav}}(P)$. 
    \textbf{(b)} By varying the power injected into the WGM cavity, we measure power-dependent transmission spectra for two different modes, each with a different eigenfrequency. 
    The WGM with the higher eigenfrequency (1047\,nm) exhibits nonlinear lineshape dependence on $P$, whereas the low eigenfrequency mode shows no power dependence. 
    The increase in transmission contrast and reduction in linewidth of the Lorentzian signifies a reduction in loss. 
    }
    \label{fig:fig3_transmissionSpectraPower}
\end{figure}

Next, we investigate the power-dependent properties of the fundamental WGMs identified in \Cref{fig:fig2_WGModes}. 
We observe that several modes exhibit power-dependent changes in their loss rates and associated transmission lineshapes, which we attribute to the excitation of a saturable absorber within the cavity. 
All investigated modes are under-coupled, accentuating the power-dependence of the cavity lineshapes.
To quantify this behaviour, we extract the internal cavity energy loss rate by measuring the fibre-taper transmission spectra across each TM WGM, focusing on TM rather than TE modes due to their greater abundance within the sampled wavelength range and to avoid any polarization-dependent effects. 
Near resonance, photons from the fibre-taper (see \Cref{fig:fig3_transmissionSpectraPower}\,(a)) couple into a WGM where they circulate until they are scattered back into the fibre-taper, absorbed by defects, or lost through other mechanisms.
By modifying the laser power, $P$, input to the fibre-taper, we measure the cavity loss rates over a range of intracavity intensity. 
Note that as in previous measurements of diamond optical cavities \cite{Mitchell:2019, Shandilya:2021, Behjat2023, Flagan:2025, Itoi2025arXiv}, we confirm that the cavity lineshape was unaffected by the use of optical amplification stages, which we operate using constant input and output power across the full measurement range (see Appendix \ref{sec:ExpSetup}).
Furthermore, in all of our experiments the laser scan speed was sufficiently low to ensure that all measurements were performed in a quasi-static regime. 
As a result, the extracted cavity loss was unaffected by the laser scan speed, within uncertainty (see Appendix \ref{sec:ExtractLoss}).

Among the modes studied, only the three modes with the highest frequencies exhibit power-dependent spectral lineshapes. In \Cref{fig:fig3_transmissionSpectraPower}~(b), we compare the transmission spectra of one of the modes that exhibits strong power dependence (1047\,nm) with that of a mode whose transmission spectra is power-independent (1604\,nm). 
The 1047\,nm mode possesses a higher intrinsic quality factor and appears as a doublet mode, while the 1604\,nm mode possesses a lower quality factor and appears as a singlet.
Note that the doublet structure arising from coherent back-scattering between clockwise and counterclockwise propagating WGMs~\cite{Borselli2005} does not directly influence defect absorption (see Appendix \ref{sec:MicrodiskCMT}).
More importantly, the 1047\,nm mode exhibits clear power-dependent changes. The transmission contrast increases with power while linewidth decreases, indicating a reduction in internal cavity loss $\kappa_{\text{cav}}$ and a corresponding increase in ${Q}$ with increasing intracavity intensity. 
In contrast, the absence of comparable power-dependent spectral changes for the 1604\,nm mode suggests that the nonlinear loss mechanism responsible for the observed behaviour is strongly wavelength-dependent. 

\begin{figure}[b!]
    \centering
    \includegraphics{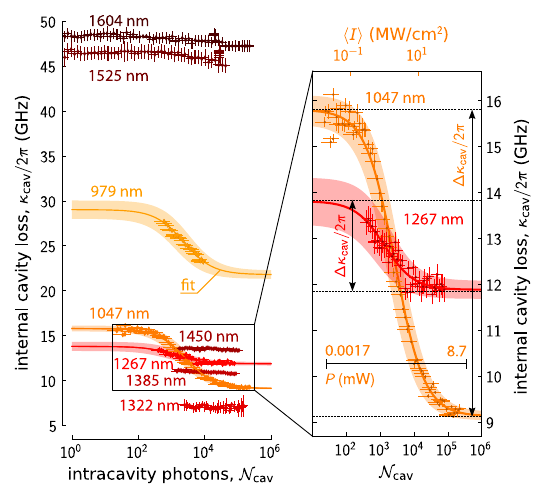}
    \caption{
    \textbf{Intensity-dependent loss in a microdisk.}
    Internal cavity loss rate as a function of intracavity photon number for WGMs between 979\,nm and 1604\,nm. 
    Three WGMs at wavelengths 979\,nm, 1047\,nm, and 1267\,nm exhibit nonlinear dependence of loss rate on optical intensity, which we attribute to a saturable absorber. 
    In the case of doublet modes, for simplicity we plot the average $\mathcal{N}_{\text{cav}}$ and $\kappa_{\text{cav}}$ of the symmetric and antisymmetric modes.
    The right plot highlights the intensity-dependent internal cavity loss rate of the 1047\,nm and 1267\,nm doublet modes, to which we fit a saturable absorber model to extract the change in internal cavity loss and saturation intensity. 
    In the right figure, the $\mathcal{N}_{\text{cav}}$ axis pertains to both wavelengths; however, the conversion between $\mathcal{N}_{\text{cav}}$ and $\langle I\rangle$ is wavelength-dependent, so the $\langle I\rangle$ axis pertains only to the 1047\,nm mode. 
    To relate the intensity and photon number to the input power, we show the range of $P$ for the 1047\,nm mode.
    The scatter points are the measured data with fit lines corresponding to \Cref{eq:fitSatAbs}.
    See the main text for more details. 
    }
    \label{fig:fig4__Saturation}
\end{figure}

To account for differences in mode properties and better understand the observed nonlinear response, we systematically investigate the intracavity intensity-dependent transmission spectra of the eight observed fundamental TM WGMs ranging from 979\,nm to 1604\,nm. The resulting transmission spectra are fit using a model derived from coupled-mode theory that incorporates coherent back-scattering~\cite{Borselli2005}, Fano interference~\cite{Suh2003}, and thermo-optic effects~\cite{Barclay2005, Itoi2025arXiv} to extract the power-dependent internal cavity loss $\kappa_{\text{cav}}$ (see Appendix \ref{sec:MicrodiskCMT})\,\cite{Itoi2025arXiv}. 
Note, however, that despite the inclusion of the thermo-optic effect, none of the modes that exhibit power-dependent lineshapes show any significant thermo-optic shifting---the inclusion of the thermo-optic effect in the fitting model serves mostly to refine our estimation of the number of intracavity photons (see Appendix \ref{sec:ExtractLoss}).
The extracted values of $\kappa_{\text{cav}}$ for each mode are plotted in \Cref{fig:fig4__Saturation} as a function of the effective intracavity photon number, $\mathcal{N}_{\text{cav}}$, which does not depend on mode type and is calculated from the power dropped into the cavity mode and the experimentally determined cavity loss rates (see Appendix \ref{sec:EigenmodeSim}).
Uncertainties in the fitted loss rates are determined from the covariance matrix obtained using the least-squares fitting procedure. These uncertainties, as well as uncertainties in the input power, $P$, and the fibre coupling efficiency, $\eta_{\text{fibre}}$, contribute to uncertainty in $\mathcal{N}_{\text{cav}}$. 
Additional details regarding the extraction of cavity loss rates are provided in Appendix \ref{sec:ExtractLoss}.

As shown in the left panel of \Cref{fig:fig4__Saturation}, significant power-dependent changes in $\kappa_{\text{cav}}$ are observed only for optical modes at 979\,nm, 1047\,nm, and 1267\,nm.
For these modes, $\kappa_{\text{cav}}$ decreases nonlinearly with increasing $\mathcal{N}_{\text{cav}}$, and, in the case of the 1047\,nm mode, clear saturation of the loss is observed. 
For clarity, the right panel of \Cref{fig:fig4__Saturation} highlights the evolution of $\kappa_{\text{cav}}$ for the 1047\,nm and 1267\,nm modes as a function of both the effective average intracavity intensity, $\langle I\rangle$, and $\mathcal{N}_{\text{cav}}$, which are related by the mode volume of each WGM.
We attribute this nonlinearity to the presence of saturable absorbers in the diamond microdisk. 
In contrast, modes at wavelengths longer than 1267\,nm exhibit no measurable intensity dependence over the investigated input power range.
Note that a wider range of input powers, and consequently a larger measurement dynamic range, is required to more conclusively determine the cavity behaviour at several of the wavelengths studied here. Of particular interest for future work is extending the power range of the 979\,nm measurements, where a significant intensity-dependent decrease in $\kappa_{\text{cav}}$ is observed but does not saturate at the highest available power setting. Further increasing the power at wavelengths between $1500$ and $1630$~nm is also of interest, as modes in this range show some indication of changes in loss at the highest available power accessible with our apparatus.

\begin{figure}[b!]
    \centering
    \includegraphics{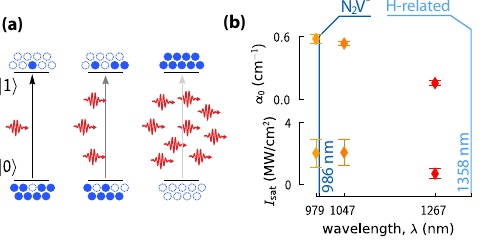}
    \caption{
    \textbf{Saturable absorption in a diamond microdisk.}
    \textbf{(a)} Saturation of defect absorbers. At low optical intensities (left), most defects are in the ground state. 
    Higher intensities (middle, right) cause the defect population to invert, reducing material absorption loss.
    \textbf{(b)} Extracted wavelength-dependent linear absorption coefficients and saturation intensities of the saturable absorbers. 
    The zero-phonon lines for the two candidate defects are shown using two vertical blue lines.}
    \label{fig:fig5_fittedSaturationValues}
\end{figure}

Applying a two-level saturable absorber model to the observed nonlinearities enables extraction of key characteristics of the defect absorbers. In a cavity, the energy loss rate due to material absorption is proportional to the absorption coefficient, $\alpha$~\cite{Saleh2019}:
\begin{align}
    \frac{\kappa_{\text{abs}}}{2\pi}= v_{\text{g}}\alpha \approx \frac{c}{n_{\text{g}}}\alpha\,,
    \label{eq:kappaAbstoAbsCoeff}
\end{align}
where $v_{\text{g}}$ and $ n_{\text{g}}$ are the group velocity and group index of the mode, respectively, $c$ is the speed of light in vacuum, and the approximation holds in the weak dispersion limit~\cite{Saleh2019}. 
Absorption by an ensemble of two-level systems is proportional to the absorber density, $M$, and the frequency-dependent absorption cross-section, $\sigma_{\omega}$, and can saturate at high intensities due to population inversion (\Cref{fig:fig5_fittedSaturationValues}~(a))\,\cite{Boyd2008}:
\begin{align}
    \alpha = \frac{M\sigma_{\omega}}{1 + \frac{\langle I\rangle}{I_{\text{sat}}}}\,,
    \label{eq:absorptionCoeff}
\end{align}
where $I_{\text{sat}}= \hbar\omega/\sigma_\omega\tau$ is the saturation intensity and $\tau$ is the lifetime of the excited state. 
Combining \Cref{eq:nonlinearCavityLoss,eq:kappaAbstoAbsCoeff,eq:absorptionCoeff} yields an intensity- and frequency-dependent expression for $\kappa_{\text{cav}}$:
\begin{align}
    \frac{\kappa_{\text{cav}}}{2\pi}= \frac{\kappa_{\text{i}}}{2\pi} + \frac{\kappa_{\text{p}}}{2\pi} + \frac{c}{n_g}\frac{\alpha_0}{1 + \frac{\langle I\rangle}{I_{\text{sat}}}}\,,
    \label{eq:fitSatAbs}
\end{align}
where $\alpha_0 = M\sigma_{\omega}$ is the linear absorption coefficient. Equation \ref{eq:fitSatAbs} is used to fit the data shown in \Cref{fig:fig4__Saturation}, yielding wavelength-dependent absorption coefficients and saturation intensities. 
Uncertainty in these extracted values incorporates uncertainty in the cavity parameters, the cavity geometry, and input power; however, the use of the cavity does not allow us to explore any polarization dependence of the absorption.
These extracted parameters, of which only those at 1047\,nm were obtained with aid from optical amplification, are presented in \Cref{fig:fig5_fittedSaturationValues}~(b) and summarized in \Cref{tab:results}.
Further details on the saturable absorber model and the extraction of parameter uncertainty can be found in Appendices \ref{sec:DeriveSatAbs} and \ref{sec:uncertainty}, respectively.

We note that the observation of saturable absorption is not limited to the cavity studied here, as we observe power-dependent cavity loss in other cavities with different radii on the same chip. We do not include those results due to the experimental and simulation overhead required to extract the saturation intensity.

\setlength{\extrarowheight}{0pt}
\begin{table}[t!]
\centering
\caption{Extracted saturation intensities and absorption coefficients at different wavelengths in diamond. 
}
\label{tab:results}
\begin{tabular}{c|c|c}
\hline \hline
wavelength, $\lambda$ (nm) & $\alpha_0$ (cm$^{-1}$) & $I_{\text{sat}}$ (MW/cm$^2$) \\
\hline 
    979                        & 0.58\,(4)                                            &2.0\,(9)                                          \\
    1047                       & 0.53\,(2)                                            & 2.1\,(8)\\
    1267                       & 0.15\,(2)                                            & 0.7\,(3)  \\
\hline
\end{tabular}
\end{table}

\section{Crystal defects and saturable absorption in diamond}
Diamond is a host to a plethora of crystal defects\,\cite{Rosa1999,Zaitsev2010,Khan2013}, many of which exhibit optical absorption.
In particular, nitrogen-\,\cite{Ashfold2020} and hydrogen-related\,\cite{Day2024} defects are common in CVD-grown diamond\,\cite{McNamara1994}.
Several of these crystal defects could contribute to the observed saturable absorption, and although we cannot conclusively determine which defect is responsible from the presented data there are a few plausible candidates.

In this work, the absorption is measured using continuous wave IR lasers---we do not use an additional high-power laser to perform pump-probe measurements\,\cite{Kehayias2013}.
The highest available laser energy is 1.27\,eV (979\,nm, see \Cref{fig:fig4__Saturation}), which, neglecting multiphoton processes, sets the upper bound for absorption. 
Only defects with absorption thresholds at or below this energy can contribute to the observed saturable absorption.
This constraint precludes several common diamond defects including substitutional nitrogen\,\cite{Rosa1999}, mono- and divacancies\,\cite{Clark1973,Pu2001,Yurgens2021}, and both charge states of the NV centre\,\cite{Shandilya2024}.

We further rule out IR absorption by the 1042~nm NV$^-$ singlet transition\,\cite{Acosta2010PRB}.
Without a sufficiently energetic pump, the NV$^-$ triplet transition ($> 1.95~\text{eV}$) will not be excited, and consequently, the singlet ground state will not be populated. 
Therefore, the NV$^-$ singlet transition is not expected to contribute appreciably to the intensity-dependent absorption observed here~\cite{Zadeh-Haghighi2024}.
Intracavity third-harmonic generation \cite{Itoi2025arXiv} could, in principle, drive NV$^-$ population to the singlet state, however this would result in a dependence of the observed absorption scaling as $\mathcal{N}_{\text{cav}}^3$, which is not observed here. 
Extending previous studies\,\cite{Shandilya2024, Itoi2025arXiv, Flagan:2025} that combine green (532 nm) and IR excitation of cavities fabricated from optical-grade samples to the dense NV cavities studied here is of significant interest. Such measurements would provide insight into the role of electronic-state dynamics of defects such as NV centres on the observed absorption. 

One plausible defect that could contribute to the observed saturable absorption is $\text{N}_2\text{V}^-$~\cite{Ashfold2020,Johnson2025,Talik2026}---an optically active defect that occurs in nitrogen-rich diamond, like the sample studied here. 
Its zero-phonon line near 986\,nm\,\cite{Johnson2025} is consistent with the intensity-dependent loss rate observed for the 979\,nm mode. However, the short lifetime ($\sim0.3$\,ns) and narrow zero-phonon line ($\sim5$\,nm) suggest that $\text{N}_2\text{V}^-$ is unlikely to significantly contribute to absorption at longer wavelengths~\cite{Johnson2025}.
A recent femtosecond Z-scan study by Talik et al. reported saturable absorption in similar defect-rich diamond material to that used here and attributed the observed nonlinearity to the
$\text{N}_2\text{V}^-$ centre\,\cite{Talik2026}. 
However, the measurement regime employed in Ref.~\cite{Talik2026} differs substantially from that used in the present work, suggesting that the two studies may probe different nonlinear absorption mechanisms or distinct defect populations.
In particular, Ref.~\cite{Talik2026} used femtosecond-pulsed excitation at intensities more than four orders of magnitude larger than those studied here and reported significantly larger absorption coefficients.
If the same absorption mechanism dominated in the nanophotonic cavities studied here, the corresponding optical loss would limit the intrinsic cavity quality factor to $Q<4000$, well below those routinely observed experimentally.
This discrepancy may arise from the different temporal regimes of the
measurements (steady-state continuous wave versus ultrafast pulsed excitation), or from additional nonlinear optical processes that become relevant only at the much higher intensities explored in Ref.~\cite{Talik2026}.
Furthermore, it is worth noting that $\text{N}_2\text{V}^0$, N$_3$VNV, and NVH centres all absorb in the visible\,\cite{Collins2005,Khan2013,Johnson2025}, allowing their contributions to be excluded. 

Another plausible candidate is a hydrogen-related defect that predominantly occurs in CVD-grown diamond\,\cite{Fuchs1995,Fuchs1995Diam,Glover:2003}.
Although neither the defect density nor its exact composition is known\,\cite{Inyushkin2023}, this defect was recently observed in a study of the same diamond material employed here\,\cite{Younesi:2025,ElementSix2021}. 
The hydrogen-related defect's zero-phonon line at 1358\,nm and broad phonon sideband~\cite{Younesi:2025} are consistent with the observed absorption at 979\,nm, 1047\,nm, and 1267\,nm, as well as its absence at wavelengths longer than 1358\,nm.
Despite it being energetically favourable, saturable absorption of the 1322\,nm mode is unexpectedly absent.
Several explanations are possible. 
First, due to several experimental restrictions, the dynamic power range at 1322\,nm was restricted and may have been insufficient to observe saturation (see Appendix\,\ref{sec:ExpSetup}). 
Second, the nonlinearity at this wavelength may be too weak to be distinguished from the measurement uncertainty. 

In summary, we do not have sufficient experimental evidence to conclusively pinpoint the defect responsible for the saturable absorption. For example, both $\text{N}_2\text{V}^-$ and the CVD-native hydrogen-related defect can contribute to the observed absorption at 979\,nm (\Cref{fig:fig5_fittedSaturationValues}\,(b)). However, for longer wavelengths, we believe that the CVD-native hydrogen-related defect\,\cite{Younesi:2025} is the most plausible candidate.
Further insight into the nature of the intensity-dependent absorption processes could be gleaned by future measurements.
Time-dependent measurements could be used to establish the saturable absorber lifetime.
Further, extensions of previous studies of nanophotonic cavities fabricated from optical and electronic-grade diamond devices, such as those in Refs.\ \cite{Mitchell:2019, 
Behjat2023, Shandilya:2021, Flagan:2025, Itoi2025arXiv}, to the wider wavelength range studied here would allow us to assess the role of defect density on intensity-dependent optical absorption. 
Similarly, creating and studying devices from high pressure high temperature (HPHT) diamond chips whose growth conditions may yield different relative concentrations of impurity and defect types than in CVD diamond could provide further insight into their effects on intensity-dependent optical absorption.

\section{Intensity-dependent loss in diamond photonic cavities}
The presence of this absorber impacts a range of applications that are sensitive to optical loss. 
In particular, these defects are detrimental to the performance of photonic cavities.
Our results indicate that the intrinsic ${Q}$ of an under-saturated, absorption-limited $(\kappa_{\text{abs}}\gg \kappa_{\text{i}}, \kappa_{\text{p}} )$ optical mode with a resonance near 1000\,nm cannot exceed $\sim5\times10^4$.
Higher $Q$ can be achieved at intracavity intensities above saturation, such as those used to measure the cavity shown in \Cref{fig:fig1_cavityLoss}\,(d).
This limitation is expected to be even more severe near the ZPLs of both defects discussed. 
These effects will primarily impact applications that rely on highly nitrogen-doped diamond, such as NV ensemble based quantum sensors.
One particularly affected application is IR absorption-based diamond magnetometry, where changes in external magnetic fields are inferred from absorption of IR (1042\,nm) light by the NV$^-$ singlet state\,\cite{Marini2008,Acosta2010,Younesi:2025}. 
Absorption by the NV$^-$ singlet state is typically inferred from a change in the transmission of a laser near 1042\,nm; however, the saturable absorbers studied here introduce additional absorption loss, thereby limiting the relative change in transmission and corresponding magnetic sensitivity\,\cite{Jensen2014, Zadeh-Haghighi2024}. 
Specifically, in a cavity-based IR absorption magnetometer the sensitivity is governed by the system's response to changes in cavity loss, and the sensitivity improves with $Q$~\cite{Dumeige2013}. Increased cavity loss from crystal defects, resulting in a lower $Q$ cavity, will negatively impact the achievable sensitivity.
However, as shown here, saturating the absorption may allow its impact to be reduced. 
The amount of power required to saturate the defect absorption is cavity geometry-dependent, but is in general inversely proportional to the effective cavity volume and the intrinsic $Q$.

Despite these limitations, the presence of a saturable absorber may provide advantages for certain photonic applications. 
Saturable absorption by solid-state defects has long been exploited for passive control of light–matter interactions, particularly in compact laser systems~\cite{Paschotta2013}. 
In microcavity platforms, an intrinsic saturable absorber can enable passive ${Q}$-switching and mode-locking, producing pulsed or modulated optical output without additional intracavity components or active modulation\,\cite{Dangoisse1990,Morris:1990}.
Although the linear absorption coefficient found here is smaller than those reported at comparable wavelengths for Cr:YAG ($\sim280$\,cm$^{-1}$) and V:YAG ($\sim1.44$\,cm$^{-1}$) saturable absorbers, diamond-based defect absorbers may nonetheless be suited to similar applications\,\cite{Dong:2001, Liu:2015}. 
Because the absorption coefficient scales with defect density (\Cref{eq:absorptionCoeff}), diamond samples with higher concentrations of the relevant defects could achieve larger saturable absorption coefficients, increasing their practical utility.
Moreover, diamond's exceptional thermal conductivity and high optical damage threshold~\cite{Shandilya2022} suggest that defect-mediated saturable absorption could support high-power or high-repetition-rate pulsed operation in diamond photonic devices.

Beyond laser systems, defect-based saturable absorbers may also enable nonlinear and all-optical signal processing in diamond.
In this work, we observe a maximum 42\,\% reduction in loss due to the saturable absorber (\Cref{fig:fig4__Saturation}), which, after accounting for the external coupling rate, corresponds to an approximately $14\,\%$ change in transmission contrast (\Cref{fig:fig3_transmissionSpectraPower}).
Although this modulation depth is partially limited by additional sources of cavity loss, improvements in fabrication that reduce background loss could significantly enhance the observable nonlinear response, increasing the utility of these defects for photonic signal processing applications.
The intensity-dependent transmission associated with saturable absorption can enable optical switching, optical logic operations\,\cite{Porzi:2008}, and is the subject of intense research within the context of neuromorphic computing \cite{Farmakidis2024}.
In nanophotonic cavities, where strong field confinement enhances light–matter interactions, even relatively weak absorbers can produce substantial nonlinear responses at relatively low input powers. 
These properties suggest that diamond microcavities incorporating such defects could serve as compact, integrable nonlinear optical elements for on-chip photonic circuits. 

\section{Conclusion}

In this work, we demonstrate and characterize saturable absorption in high-${Q}/V$ diamond microdisk cavities fabricated from dense-NV diamond.
Using power-dependent measurements of transmission spectra across whispering-gallery modes spanning 979\,nm to 1604\,nm, we observe a nonlinear reduction in cavity loss for modes near 979\,nm, 1047\,nm, and 1267\,nm.
Using a two-level saturable absorber model, we extract wavelength-dependent absorption coefficients and saturation intensities and identify a hydrogen-based defect as a plausible cause of the saturable absorption, although the $\text{N}_2\text{V}^-$ centre could also be responsible.
Absorption due to several other common defects including substitutional nitrogen, mono- and divacancies, and both charge states of the NV centre are unlikely due to energy constraints.
At 1047\,nm, we find that the defects saturate at an intensity of $2.1\,(8)\,$\,MW/cm$^2$ with an absorption coefficient of $0.53\,(2)$\,cm$^{-1}$.

These absorbers introduce a material loss mechanism that can impact the performance of diamond nanophotonic devices and is particularly relevant for quantum sensing technologies that rely on dense-NV ensembles.
One such application is IR absorption magnetometry.
Here, we characterize an optical mode (${Q}>7\times10^4$) overlapping with the 1042 nm singlet transition of the negatively charged NV, demonstrating that these microcavities remain promising candidates for IR absorption magnetometry despite the presence of additional defect-mediated loss. 
Simultaneously, the observed nonlinear response highlights the potential of intrinsic diamond defects as functional elements for nonlinear and all-optical photonic applications, such as ${Q}$-switching, logic, and neuromorphic computing operations. 
These results provide new insight into defect-mediated optical loss in diamond nanophotonics and suggest strategies to harness defect-induced nonlinearities in future diamond photonic devices.
\section*{Funding}
This work was supported by the Natural Science and Engineering Research Council (Discovery Grant, Research Tools and Instruments, Alliance Quantum ALLRP 586298-2023, the CanQuEST Alliance Quantum Consortia, and the ARAQNE Alliance Quantum Consortia), the NSERC Alliance - Alberta Innovates Advance Program, and the Swiss National Science Foundation (P500PT_206919).

\section*{Acknowledgments}
We acknowledge and thank Joe Itoi and Vinaya K. Kavatamane for helpful discussions. 

\section*{Disclosures}
The authors declare no conflicts of interest.

\section*{Data availability}
Experimental data underlying the results presented in this paper are not publicly available at this time but may be obtained from the authors upon reasonable request.

\newpage
\appendix

\section{Coupled mode theory}
\label{sec:MicrodiskCMT}
\begin{figure*}[b!]
    \centering
    \includegraphics[width = \textwidth]{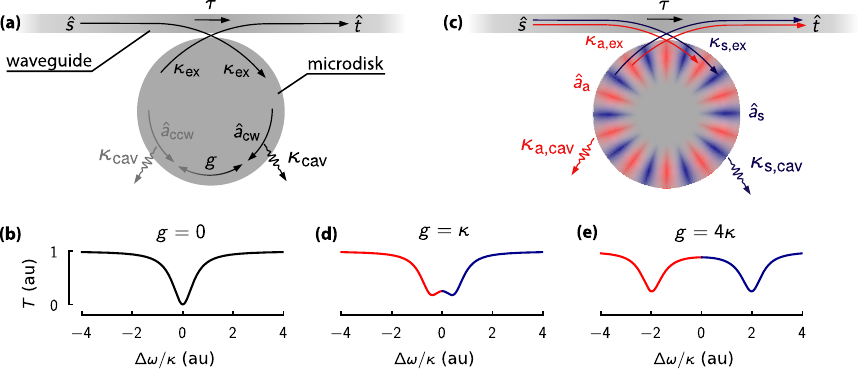}
    \caption{\textbf{Coupled mode theory for a WGM resonator.} 
    \textbf{(a)} An ideal WGM resonator supports identical degenerate modes, one travelling clock-wise (CW) and the other travelling counter-clock-wise (CCW). 
    \textbf{(b)} The transmission spectrum of a coupled waveguide takes the form of a single Lorentzian.
    In the case that coherent scattering between the two modes is non-negligible ($g\sim\kappa$) the degeneracy is lifted and two standing-wave modes form in the WGM resonator \textbf{(c)}: one symmetric (blue) and one antisymmetric (red). The nodes of one mode are located at the anti-nodes of the other.
    Depending on the strength of the coupling, $g$, the transmission spectrum takes the form of a coherent sum of two offset Lorentzians \textbf{(d, e)}. 
    }
    \label{fig:fig6_CMT_WGM_resonators}
\end{figure*}
To extract the defect-induced loss rate from the cavity transmission scans, we use coupled mode theory to model the cavity dynamics. 
Specifically, coupled mode or input-output theory gives us an equation to describe the dynamics of the field amplitudes inside the cavity~\cite{Haus1984}. 

For a given azimuthal number, $m$, an ideal WGM resonator supports identical degenerate modes---one travelling clock-wise (CW) and the other travelling counter-clock-wise (CCW), as shown in \Cref{fig:fig6_CMT_WGM_resonators}\,a. 
The waveguide (in our case, a single-mode fibre-taper) carries an input field $\hat{s}$, which couples to the CW cavity mode $\hat{a}_{\text{cw}}$ with energy coupling rate $\kappa_{\text{ex}}$, which then couples back out into the fibre.
The amplitude of the CW cavity mode depends not only on the rate of external coupling, but also on internal cavity loss rates.
Fabrication imperfections in the cavity and surface scattering introduce intrinsic loss ($\kappa_{\text{i}}$), the presence of the fibre-taper introduces parasitic loss from coupling to high-order fibre modes or radiation modes ($\kappa_{\text{p}}$), and the presence of atomic defects introduces absorption losses ($\kappa_{\text{abs}}$). 
The total cavity energy loss rate for the CW mode can then be written as~\cite{Haus1984} 
\begin{align}
    \kappa = \kappa_{\text{ex}} +  \kappa_{\text{i}}+ \kappa_{\text{p}}+ \kappa_{\text{abs}}\,.
\end{align}
It is typically a good approximation to assume that this loss rate identically describes the CCW mode, as sketched in \Cref{fig:fig6_CMT_WGM_resonators}\,a.
For simplicity's sake we also define an internal cavity loss rate, 
\begin{align}
    \kappa_{\text{cav}} \equiv \kappa_{\text{i}} + \kappa_{\text{p}} + \kappa_{\text{abs}}\,.
    \label{eq:kappaInt}
\end{align}

The above definitions allow us to write down an equation of motion used to describe the cavity field amplitude~\cite{Haus1984,Barclay2009OpticsExpress}
\begin{align}
    \dot{a}_{\text{cw}} = \sqrt{\kappa_{\text{ex}}}\hat{s} + i\Delta\omega \hat{a}_{\text{cw}} - \frac{\kappa}{2} \hat{a}_{\text{cw}}\,,
    \label{eq:EoM-cw_ccw_modes}
\end{align}
where $\Delta\omega = \omega -\omega_0$ describes the cavity frequency detuning between the laser field at frequency $\omega$ and the cavity resonance at frequency $\omega_0$. 
The term $\kappa/2$ denotes the field amplitude decay rate of the cavity, which contains $\mathcal{N} = \langle \hat{a}_{\text{cw}} \rangle^2$ intracavity photons, given an input power of $P = \hbar \omega \langle \hat{s} \rangle^2$. 

In the steady-state approximation, the amplitude of the cavity field is 
\begin{align}
    \langle\hat{a}_{\text{cw}}\rangle = \frac{\sqrt{\kappa_{\text{ex}}} }{\frac{\kappa}{2} - i\Delta\omega}\langle\hat{s}\rangle\,.
    \label{eq:intracavityField1}
\end{align}
Input-output theory\,\cite{Haus1984,Clerk2010PhysRev} then lets us write down the transmitted field, $\langle\hat{t}\rangle = \tau \langle\hat{s}\rangle - \sqrt{\kappa_{\text{e}}}\langle\hat{a}_{\text{cw}}\rangle$, followed by the normalized cavity transmission intensity:
\begin{align}
    T \equiv \left| \frac{\langle\hat{t}\rangle}{\langle\hat{s}\rangle}\right|^2 = \left| \tau - \frac{\kappa_{\text{ex}}}{\frac{\kappa}{2}-i\Delta\omega}\right|^2\,.
    \label{eq:transmission1}
\end{align}
The transmission coefficient $\tau$ is a complex number and can result in Fano asymmetries in the transmission spectra due to interference between the input and cavity fields\,\cite{Suh2003,Itoi2025arXiv}. 
An example transmission spectrum for a critically coupled cavity mode, i.e. $\kappa_{\text{cav}}=\kappa_{\text{ex}}$, is shown in \Cref{fig:fig6_CMT_WGM_resonators}\,(b).

\subsection{Doublet modes}
\label{sec:doublets}
The microdisk supports degenerate WGM modes---one travelling in the clockwise (CW) direction and the other travelling in the counterclockwise (CCW) direction. Surface roughness and imperfections can lead to coherent Rayleigh scattering between the CW and CCW modes\,\cite{Gorodetsky2000,Lake2016}. When the back-scattering rate $g$ is non-negligible, i.e. $g\sim \kappa$, the degeneracy between the CW and CCW modes is lifted, resulting in the formation of orthogonal standing-wave modes (\Cref{fig:fig6_CMT_WGM_resonators}~(c)) whose splitting is set by the back-scattering rate (\Cref{fig:fig6_CMT_WGM_resonators}~(d,e))\,\cite{Borselli2004,Borselli2005}. By incorporating this back-scattering, the modified equation of motion (\Cref{eq:EoM-cw_ccw_modes}) becomes

\begin{align}
    \dot{\hat{a}}_{\text{cw}} &=  \left(i\Delta\omega - \frac{\kappa}{2}\right) \hat{a}_{\text{cw}} + i\frac{g}{2}\hat{a}_{\text{ccw}} +\sqrt{\kappa_{\text{ex}}}\hat{s}\,, \label{eq:EoM-coupling-1}\\
     \dot{\hat{a}}_{\text{ccw}} &= \left(i\Delta\omega - \frac{\kappa}{2}\right) \hat{a}_{\text{ccw}} + i\frac{g}{2}\hat{a}_{\text{cw}}\,.
     \label{eq:EoM-coupling-2}
\end{align}
These equations are most easily solved in a new \textit{standing mode} basis\,\cite{Borselli2006}
\begin{align}
    \hat{a}_{\text{s}} &= \frac{1}{\sqrt{2}}\left(\hat{a}_{\text{cw}} + \hat{a}_{\text{ccw}}\right)\,, \label{eq:cw_to_standing_wave-1}\\
    \hat{a}_{\text{a}} &= \frac{1}{\sqrt{2}}\left(\hat{a}_{\text{cw}} - \hat{a}_{\text{ccw}}\right)\,,
    \label{eq:cw_to_standing_wave-2}
\end{align}
where $\hat{a}_{\text{s}}$ and $\hat{a}_{\text{a}}$ describe the symmetric and antisymmetric standing-wave modes in the WGM resonator. 
We then use \Cref{eq:EoM-coupling-1,eq:EoM-coupling-2,eq:cw_to_standing_wave-1,eq:cw_to_standing_wave-2} to produce two new equations of motion:
\begin{align}
    \dot{a}_{\text{s}} &= \left[i\left(\Delta\omega-\frac{g}{2}\right) - \frac{\kappa}{2}\right] \hat{a}_{\text{s}} + \sqrt{\frac{\kappa_{\text{ex}}}{2}}\hat{s}\,, 
         \label{eq:EoM-new-1}\\
     \dot{a}_{\text{a}} &= \left[i\left(\Delta\omega+\frac{g}{2}\right) - \frac{\kappa}{2}\right] \hat{a}_{\text{a}} + \sqrt{\frac{\kappa_{\text{ex}}}{2}}\hat{s}\,.
     \label{eq:EoM-new-2}
\end{align}

Heuristically, these equations imply that the resonance frequencies of the two modes are split by $g = \omega_a - \omega_s$.
They also imply that the input field couples equally well into $\hat{a}_{\text{s}}$ and $\hat{a}_{\text{a}}$; however, that is not necessarily the case, as the field overlap between the waveguide and $\hat{a}_{\text{s}}$ or $\hat{a}_{\text{a}}$ depends on the position of the waveguide with respect to each mode. 
Further, the internal loss rates associated with each mode can be different, depending on how the modes interacts with the imperfections at the microdisk surface\,\cite{Masuda:2024}.
To generalize \Cref{eq:EoM-new-1,eq:EoM-new-2} to account for these differences, we introduce standing-mode-dependent loss rates ($\kappa_{{j}}= \kappa_{j,\text{ex}}+\kappa_{j,\text{in}}$ for $j = \text{s}, \text{a}$), which allow us to express the steady-state field amplitudes as
\begin{align}
    \langle\hat{a}_{\text{s}}\rangle = \frac{\sqrt{\frac{\kappa_{s,\text{ex}}}{2}} }{\frac{\kappa_{\text{s}}}{2} - i\left(\Delta\omega-\frac{g}{2}\right)}\langle\hat{s}\rangle\,, \label{eq:intracavityFieldSymm} \\
    \langle\hat{a}_{\text{a}}\rangle = \frac{\sqrt{\frac{\kappa_{a,\text{ex}}}{2}} }{\frac{\kappa_{\text{a}}}{2} - i\left(\Delta\omega+\frac{g}{2}\right)}\langle\hat{s}\rangle\,    \label{eq:intracavityFieldAsymm}.
\end{align}
The normalized transmission intensity is then
\begin{align}
    T &= \left| \tau -\sqrt{ \frac{\kappa_{s,\text{ex}}}{2}}\frac{\langle\hat{a}_{\text{s}}\rangle}{\langle\hat{s}\rangle} 
    - \sqrt{ \frac{\kappa_{a,\text{ex}}}{2}}\frac{\langle\hat{a}_{\text{a}}\rangle}{\langle\hat{s}\rangle}\right|^2\\
    &= \left| \tau - \frac{\kappa_{s,\text{ex}}}{2}\frac{1}{\frac{\kappa_{\text{s}}}{2} - i\left(\Delta\omega-\frac{g}{2}\right)} - \frac{\kappa_{a,\text{ex}}}{2}\frac{1}{\frac{\kappa_{\text{a}}}{2} - i\left(\Delta\omega+\frac{g}{2}\right)}\right|^2\,.
    \label{eq:transmission2}
\end{align}

\subsection{Thermo-optic effects}
Populating the optical cavity with large photon numbers causes heating, as dictated by thermo-optic effects~\cite{Barclay2005}. 
Heating leads to thermal expansion (i.e. the cavity gets larger), which red-shifts the modes. 
In addition, heating modifies the refractive index via the thermo-optic effect.
This effect can be modelled by making the cavity detuning parameter, $\Delta\omega$, intracavity-field-dependent\,\cite{Itoi2025arXiv}. 
Specifically, replacing $\Delta\omega \rightarrow \Delta\omega - c_{\text{T}}\langle\hat{a}\rangle^2$, \Cref{eq:intracavityFieldSymm,eq:intracavityFieldAsymm} become
\begin{align}
     \langle\hat{a}_{j}\rangle = \frac{\sqrt{\frac{\kappa_{\text{ex,}j}}{2}} }{\frac{\kappa_{j}}{2} - i\left(\Delta\omega \pm\frac{g}{2} -c_{\text{T}}\langle\hat{a}_{j}\rangle^2\right)}\langle\hat{s}\rangle\label{eq:intracavityFieldSymmTO}\,
\end{align}
for $j = \text{s}, \text{a}$. Here, $c_{\text{T}}$ is the thermo-optic coefficent, which depends on the geometry and material composition. 
Equation \ref{eq:intracavityFieldSymmTO} is cubic in nature, meaning that there are three cavity field solutions; however, we
only consider the real part of the solution that matches the experimental results\,\cite{Itoi2025arXiv}---generally, this corresponds with the solution that gives the fewest intracavity photons, $\mathcal{N}_j =  \langle\hat{a}_j\rangle^2 $.
The normalized intensity transmission (\Cref{eq:transmission2}) becomes
\begin{align}
\begin{aligned}
T = \Bigg| \tau 
&- \frac{\kappa_{s,\text{ex}}}{2}
   \frac{1}{\frac{\kappa_{\text{s}}}{2} - i\left(\Delta\omega-\frac{g}{2}- c_{\text{T}}\mathcal{N}_{\text{s}}\right)} \\
&- \frac{\kappa_{a,\text{ex}}}{2}
   \frac{1}{\frac{\kappa_{\text{a}}}{2} - i\left(\Delta\omega+\frac{g}{2}- c_{\text{T}}\mathcal{N}_{\text{a}}\right)}\Bigg|^2.
\end{aligned}
\label{eq:transmissionDoubletTO}
\end{align}
Note that mode splitting is governed by the local variation in the real part of the cavity's dielectric constant, for example from surface roughness, and its overlap with a given optical mode's field. Although a given mode splitting is more apparent for high-$Q$ resonances, it is not affected by changes in internal loss due to intensity-dependent optical absorption.

\section{Experimental setup}
\label{sec:ExpSetup}
\begin{figure*}[ht!]
    \centering
    \includegraphics[width = \textwidth]{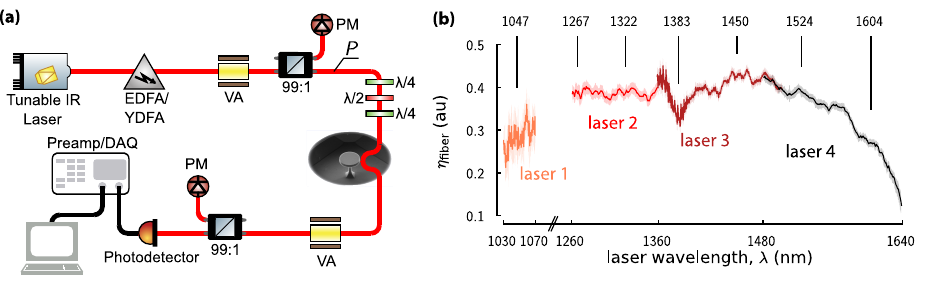}
    \caption{
    \textbf{Experimental setup and spectral fibre transmission efficiency.}
    \textbf{(a)} The experiment uses a range of different tunable IR lasers, whose power is controlled using fibre amplifiers (EDFA/YDFA) and variable attenuators (VA). 
    The transmitted laser light is measured using a power meter (PM) before being injected into a microdisk using a dimpled fibre-taper. 
    The transmission is monitored on a photodetector. 
    See the main text for more detail. 
    \textbf{(b)} The bottom axis shows the combined wavelength range of four of the lasers, and the top axis shows the wavelength of each TM mode investigated in this manuscript.
    Each lineplot shows an averaged transmission spectrum, whereas the shaded region around each lineplot shows the range of each measurement or the effective uncertainty. The 
    transmission scan of the TLB-6719 (940-985 nm) laser is not shown.}
    \label{fig:fig7_fibreEfficiency}
\end{figure*}

The sample under investigation is a  ``quantum-grade" diamond with high NV density ($[\text{NV}]=4.5\,$ppm), grown by Element Six (DNV$^{\text{TM}}$ B14 sample type) using chemical vapour deposition (CVD)\,\cite{ElementSix2021}.
Diamond microdisks are fabricated from the substrate using a quasi-isotropic reactive ion etch undercut method~\cite{Khanaliloo:2015,Mitchell:2019}.
The resulting devices support WGMs, and these same devices have been used to perform fluorescence-based magnetometry\,\cite{Sorensen:2025}.

Coherent mode spectroscopy\,\cite{Masuda:2024} for wavelengths between  940---1640\,nm is performed to characterize the microdisk using the experimental setup is shown schematically in \Cref{fig:fig7_fibreEfficiency}\,(a). 
To span the wavelength range, we use five different lasers including two Newport lasers (TLB-6719, $940-985$\,nm and TLB-6721, $1030-1070$\,nm), and three Santec lasers (TSL-570 $1260-1360$\,nm, $1357-1503$\,nm, $1480-1640$\,nm). 
To extend the dynamic intensity range of some of the transmission measurements, we optically amplify the transmitted light of the TLB-6721 ($1030-1070$\,nm) and TSL-570 ($1480-1640$\,nm) lasers using an yttrium-doped fibre amplifier (Thorlabs YDFA100S) and an erbium-doped fibre amplifier (Pritel LNHP-FA-27-IO-CP), respectively.  
Note that for all input power levels, broadband amplified spontaneous emission from the amplifiers is weak compared to the amplified laser field.
Prior to cavity insertion, an EXFO FVA-3100 variable attenuator is used to control the input power. 
Fibre coupling to a diamond microdisk is achieved using a dimpled fibre-taper~\cite{Michael2007, Masuda:2024} and coupling is maximized by optimizing polarization with the use of a fibre paddle controller. The polarization was monitored and observed to be stable throughout the measurements.
The cavity output couples back into the dimpled fibre-taper and is further attenuated by a second EXFO FVA-3100 in order to maintain approximately constant incident power on the New Focus nanosecond IR photodetector (Model 1621 or 1623 as required by probe wavelengths being measured).

Several extracted parameters, including the intracavity photon number (\Cref{eq:intracavityPhotonNumber}), depend on the transmission efficiency of the fibre, $\eta_{\text{fibre}}$, at different wavelengths. 
Therefore, for each laser we directly measure $\eta_{\text{fibre}}$, as shown in \Cref{fig:fig7_fibreEfficiency}\,(b). 
Note that the off-resonance fibre transmission was not observed to vary with changes in power, indicating that the fibre-taper optical properties are independent of input power.

\section{Cavity parameters and eigenmode simulations}
\label{sec:EigenmodeSim}
\begin{figure*}[t!]
    \centering
    \includegraphics{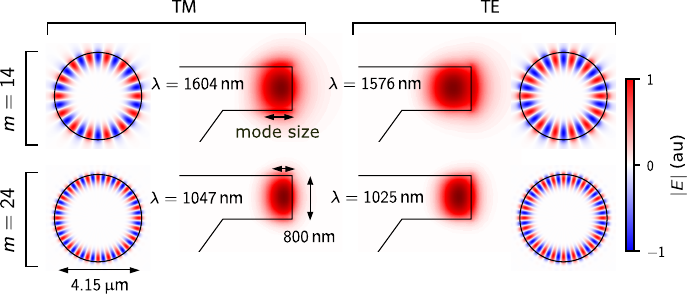}
    \caption{\textbf{Eigenmode simulations. } Spatial field profiles of the diamond microdisk WGMs. The electric field profile of four modes---two TM modes and two TE modes---is shown from from two profiles: from the top, and from a cross-section. 
    The measured TM modes align very well with the simulated TM modes, as demonstrated by \Cref{fig:fig2_WGModes}~(b)}.
    
    \label{fig:fig8_fieldSimulation}
\end{figure*}

Studying the intensity- and wavelength-dependence of absorption by defects necessitates knowledge of the 
spatial distribution of the optical modes.
To do so, we simulate the optical cavity mode using the COMSOL Multiphysics finite element solver, adjusting the radius and thickness of the simulated microdisk so that the resulting eigenfrequencies of the simulated fundamental cavity modes matched the measured ones\,\cite{Behjat2023}. 
This approach gives us a better estimate of the cavity geometry.
Specifically, the best fit corresponds to a microdisk diameter of $d_{\text{disk}}=4.15\,\upmu$m and a thickness of $t_{\text{disk}}=800\,\upmu$m. 
Note that these simulations account for material dispersion, assumed to be the same as bulk diamond\,\cite{Phillip1964}.
Four of the simulated electric field mode profiles are shown in \Cref{fig:fig8_fieldSimulation}---specifically, we plot the TM and TE fundamental modes for azimuthal numbers $m = 14, 24$. 
The figure highlights how higher $m$-number modes are more confined within the diamond resonator and have smaller effective mode volumes than smaller $m$-number modes, as demonstrated by the \textit{mode size} annotation. 
It also highlights that the TM modes are more confined towards the outside of the disk than their TE-mode counterparts. 

In the following sub-sections we elucidate the calculation of various cavity parameters obtained from simulations. 
These cavity parameters are used in concert with the measurement-inferred intracavity photon number to deduce the intracavity optical intensity. 

\subsection{Average cavity intensity}
\label{sec:avgCavityIntensity}
When considering a nonlinear optical absorption process within a nanophotonic device, it is often desirable to spatially average the local strength of the process over the cavity mode energy density distribution and, in this way, derive an effective mode volume relating intracavity energy to absorption rate \cite{Barclay2005}. In the case of saturable absorption, this approach is complicated by the non-polynomial scaling of absorption with intensity. 
We instead approximate the average intensity following the process described below. 

The time-averaged intensity of an optical field with energy density $u(\mathbf{r})$ at position $\mathbf{r}$ is given by
\begin{align}
    I(\mathbf{r}) = v_{\text{g}}u(\mathbf{r}) \,,
    \label{eq:Intensity}
\end{align}
where, $v_{\text{g}} = c/n_{\text{g}}$ is the group velocity, $c$ is the speed of light in vacuum, and $n_{\text{g}}$ is the group index. 
In this work we study single-photon absorption processes, for which the average intensity inside the diamond is the relevant quantity. 
Assuming that the radial field profile of the mode can be approximated by a Gaussian and that most of the optical field is confined within the diamond absorbing material, we compute a power-weighted average intensity:
\begin{align}
    \langle I\rangle &= \frac{\int I^2(\mathbf{r})\text{d}A}{\int I(\mathbf{r})\text{d}A}\\
    &= \frac{1}{2}\mathrm{max}[I(\mathbf{r})].
    \label{eq:averageIntensity}
\end{align}
To justify this approximation we evaluate the confinement factor $\Gamma_0$, which gives the fraction of the total electromagnetic energy contained inside the diamond~\cite{Barclay2007}:
\begin{align}
    \Gamma_0 = \frac{\int_{\text{diamond}} u(\mathbf{r})\, d^3 \mathbf{r}}{\int u(\mathbf{r})\, d^3 \mathbf{r}}\,.
    \label{eq:confinementFactor}
\end{align}
This confinement factor should be near-unity for the assumption to be valid.
To facilitate the calculation of \Cref{eq:averageIntensity}, we define two additional quantities: the effective mode volume $V_{\text{eff}}$ and the intracavity photon number $\mathcal{N}_{\text{cav}}$:
\begin{align}
    V_{\mathrm{eff}} &=
    \frac{\int u(\mathbf{r})\, d^3\mathbf{r}}
    {\max \left[u(\mathbf{r})\right]}\,,
    \label{eq:effectiveModeVolume}\\
    \mathcal{N}_{\mathrm{cav}} &=
    \frac{1}{\hbar \omega}
    \int u(\mathbf{r})\, d^3 \mathbf{r}\,.
    \label{eq:Ncav}
\end{align}
The effective mode volume is often expressed in terms of the cavity wavelength $\lambda$, yielding a dimensionless mode volume
\begin{align}
    V_0 =
    V_{\text{eff}}
    \left(\frac{\lambda}{n_{\text{eff}}}\right)^{-3},
    \label{eq:dimensionlessModeVolume}
\end{align}
where $n_{\text{eff}}$ is the effective refractive index of the cavity mode.

Combining \Cref{eq:averageIntensity,eq:effectiveModeVolume,eq:Ncav} yields
\begin{align}
    \langle I\rangle =
    \frac{c\mathcal{N}_{\text{cav}}\hbar\omega}{2n_{\text{g}}V_{\text{eff}}}.
    \label{eq:averageIntensity2}
\end{align}
The validity of \Cref{eq:averageIntensity2} can be assessed by evaluating $\Gamma_0$ using \Cref{eq:confinementFactor} together with mode simulations.
To evaluate \Cref{eq:averageIntensity2}, several parameters must be determined. Both $n_{\text{g}}$ (see \Cref{sec:groupIndex}) and $V_{\text{eff}}$ can be extracted from numerical simulations, while $\mathcal{N}_{\text{cav}}$ can be calculated from the experimental parameters using
\begin{align}
    \mathcal{N}_{\text{cav}} &=
    \frac{P\sqrt{\eta_{\text{fibre}}}}{\hbar \omega}
    \left(\frac{\langle\hat{a}\rangle}{\langle\hat{s}\rangle}\right)^2.
    \label{eq:intracavityPhotonNumber}
\end{align}
Here, $P$ is the optical power injected into the fibre-taper and $\eta_{\text{fibre}}$ is the transmission efficiency of the fibre-taper at frequency $\omega$.

In the case of a cavity doublet, the number of intracavity photons on resonance ($\Delta\omega \pm g/2 - c_{\text{T}}\mathcal{N}_{\text{cav}}=0$) can be calculated for each standing-wave mode:
\begin{align}
    \mathcal{N}_{\text{cav}}
    =
    \frac{P\sqrt{\eta_{\text{fibre}}}}{\hbar \omega}
    \frac{2\kappa_{\text{ex},j}}{\left(\kappa_{\text{ex},j}+\kappa_{\text{c},j}\right)^2 },
\end{align}
for $j\in\{\text{s},\text{a}\}$.
When the thermo-optic shift is non-negligible, \Cref{eq:intracavityPhotonNumber} becomes cubic and yields three complex solutions. 
In such cases we retain the solution whose real part is consistent with the experimentally observed intracavity photon number.

For a cavity singlet, only one travelling-wave mode is populated. To further simplify \Cref{eq:intracavityPhotonNumber}, we consider a singlet cavity mode ($g=0$) on resonance ($\Delta\omega=0$) with no thermo-optic shift ($c_{\text{T}}=0$). Under these conditions, the intracavity photon number reduces to
\begin{align}
    \mathcal{N}_{\text{cav}}(g=\Delta\omega=0)
    =
    \frac{P\sqrt{\eta_{\text{fibre}}}}{\hbar \omega}
    \frac{4\kappa_{\text{ex}}}{\left(\kappa_{\text{ex}}+\kappa_{\text{cav}}\right)^2}.
\end{align}
Accurate determination of the saturation intensity therefore requires a reliable estimate of $\eta_{\text{fibre}}$, which is determined in Appendix \ref{sec:ExpSetup}.

\subsection{Group index}
\label{sec:groupIndex}
Next, we discuss the group index of the resonator, which is used to convert the intracavity photon number into intensity and can be calculated using\,\cite{Yariv2007}
\begin{align}
    n_{\text{g}} = c\frac{\partial k_m}{\partial \omega_m}\,.
    \label{eq:groupIndex1}
\end{align}
The group index differs from the effective refractive index, which is approximated by
\begin{align}
    n_{\text{eff}} = \frac{cm}{\omega R_{\text{eff}}}\,.
    \label{eq:neff}
\end{align}
Here, $k_m = m/R_{\text{eff}}$ and $\omega_m$ are the wavenumber and frequency, respectively, of the $m$-th mode, with effective radius $R_{\text{eff}}$. 
The effective radius is calculated from the following weighted volume integral:
\begin{align}
   R_{\text{eff}} = \frac{\int r\epsilon(\mathbf{r})|\mathbf{E}(\mathbf{r})|^2d^3\mathbf{r}}{\int\epsilon(\mathbf{r})|\mathbf{E}(\mathbf{r})|^2d^3\mathbf{r}}\,,
   \label{eq:Reff}
\end{align}
where $r$ is the axial distance from the centre of the rotationally symmetric microdisk. 
Generally, both $m$ and $R_{\text{eff}}$ are dispersive, so \Cref{eq:groupIndex1} becomes
\begin{align}
    n_{\text{g}} = \frac{c}{R_{\text{eff}}\frac{\partial\omega}{\partial m}} - \frac{cm}{R_{\text{eff}}^2}\frac{\partial R_{\text{eff}}}{\partial \omega}\,.
    \label{eq:groupIndex2}
\end{align}
All parameters in \Cref{eq:groupIndex2} can be deduced from simulations of the cavity modes performed using COMSOL Multiphysics.

\setlength{\extrarowheight}{0pt}
\begin{table*}[b!]
\centering
\caption{Device parameters for different TM optical modes, calculated from simulation. 
}
\label{tab:TMModeParameters}
\begin{tabular}{ccccccccc}
\hline \hline
$m$ & $\lambda_{\text{meas}}$ (nm) & $\lambda_{\text{cav}}$ (nm) & $\omega_{\text{cav}}/2\pi$ (THz) & $V_0$ & $\Gamma_0$ & $R_{\text{eff}}$ ($\upmu$m) & $n_{\text{eff}}$ & $n_{\text{g}}$ \\ \hline
14  & 1604                         & 1605                        & 186.8                            & 7.86                                                                      & 0.960      & 1.84                        & 1.94             & 2.50           \\
15  & 1524                         & 1524                        & 196.7                            & 8.64                                                                      & 0.965      & 1.85                        & 1.97             & 2.49           \\
16  & 1450                         & 1451                        & 206.6                            & 9.49                                                                      & 0.969      & 1.85                        & 1.99             & 2.47           \\
17  & 1383                         & 1385                        & 216.5                            & 10.4                                                                      & 0.972      & 1.86                        & 2.02             & 2.46           \\
18  & 1322                         & 1324                        & 226.4                            & 11.4                                                                      & 0.974      & 1.86                        & 2.03             & 2.45           \\
19  & 1267                         & 1269                        & 236.3                            & 12.4                                                                      & 0.977      & 1.87                        & 2.05             & 2.44           \\
20  & ---                           & 1217                        & 246.3                            & 13.4                                                                      & 0.979      & 1.87                        & 2.07             & 2.43           \\
21  & ---                           & 1170                        & 256.3                            & 14.5                                                                      & 0.980      & 1.88                        & 2.08             & 2.43           \\
22  & ---                           & 1126                        & 266.3                            & 15.7                                                                      & 0.982      & 1.88                        & 2.10             & 2.42           \\
23  & ---                           & 1085                        & 276.2                            & 17.0                                                                      & 0.983      & 1.89                        & 2.11             & 2.42           \\
24  & 1047                         & 1047                        & 286.2                            & 18.4                                                                      & 0.984      & 1.89                        & 2.12             & 2.41           \\
25  & ---                           & 1012                        & 296.2                            & 19.8                                                                      & 0.985      & 1.89                        & 2.13             & 2.41           \\
26  & 979                          & 979                         & 306.2                            & 21.3                                                                      & 0.986      & 1.90                        & 2.14             & 2.41           \\ \hline
\end{tabular}
\end{table*}

\subsection{Simulation results}
We now simulate the spatial mode profile for all relevant optical modes. 
The simulated eigenfrequency, $\omega_{\text{cav}}$, and azimuthal number, $m$, of each fundamental mode are plotted in \Cref{fig:fig8_fieldSimulation}\,(b) alongside their experimentally measured counterparts. 
The proximity of one of the TM modes to the 1042\,nm NV singlet transition and the greater number of TM modes within the sampled wavelength range motivate the study of their properties as opposed to those of the TE-mode counterparts.  
We must also note that the absence of measured data at TM-mode eigenfrequencies corresponding to $m\in\{20, 21, 22, 23, 25\}$ is due to the lack of a laser at those wavelengths. 
By simulating the spatial distribution of each mode, we can ascertain the effective mode volume, refractive index, and group index, allowing us to determine the photon-number-dependent intensities at different wavelengths.
We summarize these parameters as predicted by simulation for the TM modes in \Cref{tab:TMModeParameters}, calculated using \Cref{eq:dimensionlessModeVolume,eq:confinementFactor,eq:Reff,eq:neff,eq:groupIndex2}.
The near-unity values of the $\Gamma_0$ in \Cref{tab:TMModeParameters} validate the assumption that most of the field is contained by the diamond, thereby substantiating \Cref{eq:averageIntensity2}.
Note that while $\mathcal{N}_{\text{cav}}$ is agnostic to mode identification, the conversion of $\mathcal{N}_{\text{cav}}$ to $\langle I\rangle$ is reliant upon the accurate assignment of each measured mode to a given simulated mode. 
Table \ref{tab:TMModeParameters} shows that mode volume varies between each mode studied here. Errors in mode assignment will lead to errors in calculating $\langle I\rangle$ from $\mathcal{N}_{\text{cav}}$, however, uncertainties in $\langle I\rangle$ from $\mathcal{N}_{\text{cav}}$ are dominated by contributions from the measured input power, fibre transmission efficiency, and selection of the effective $\mathcal{N}_{\text{cav}}$. 
The effects of errors in group index and effective mode volume on the extracted saturation intensity and absorption coefficients are minimal compared to other sources of uncertainty. 
See the discussion in Appendix \ref{sec:uncertainty}.

\section{Extraction of cavity loss rates}
\label{sec:ExtractLoss}
\begin{figure*}[t!]
    \centering
    \includegraphics[width=\textwidth]{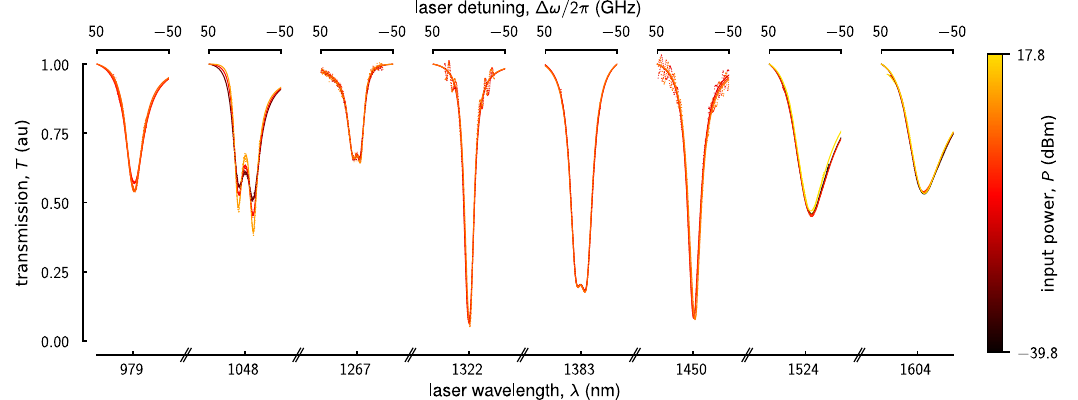}
    \caption{
    \textbf{Power-dependent cavity transmission spectra}. 
    The transmission spectra of several fundamental TM WGMs measured at different input powers are fit (\Cref{eq:transmissionDoubletTO}) and plotted. 
    Loss parameters and intracavity intensity are extracted from each of these transmission spectra.
    }
    \label{fig:fig9_SI_dataTransmissionKappas}
\end{figure*}

\begin{figure}[t!]
    \centering
    \includegraphics{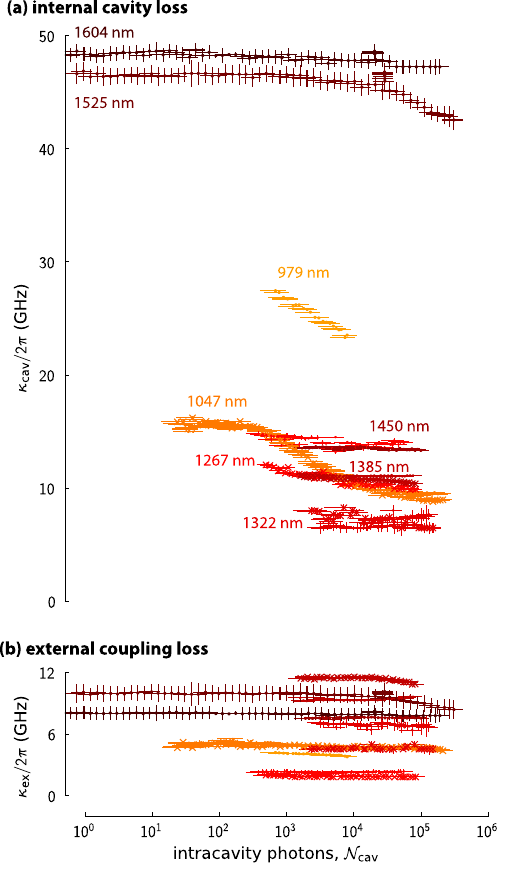}
    \caption{
    \textbf{Power-dependent cavity loss}. We extract power-dependent cavity- and external-loss rates ($\kappa_{\text{cav}}$ and $\kappa_{\text{ex}}$) from the transmission spectra shown in \Cref{fig:fig9_SI_dataTransmissionKappas} and plot them in \textbf{(a)} and \textbf{(b)}, respectively.
    For doublet modes, we plot the cavity loss rates for the symmetric and antisymmetric modes separately and denote them using dot and cross scatter points, respectively. 
    The internal cavity loss rates for most modes remain relatively constant, with the exception of those at 979\,nm, 1047\,nm, and 1267\,nm whereas the external coupling rates remain nearly constant for all modes. 
    The change in $\kappa_{\text{cav}}$ at high powers of the 1525\,nm mode correspond with considerable changes in $\kappa_{\text{ex}}$ signifying either a change in fibre position or an unreliable fit. 
    We omit this data in \Cref{fig:fig4__Saturation}.
    Uncertainties in the fitted values were estimated from the residuals of the Jacobian matrix obtained in the least-squares fit, whereas uncertainty in intracavity photon number is primarily due to uncertainty in the measurements of $P$ and $\eta_{\text{fibre}}$. 
    }
    \label{fig:fig10_SI_dataKappas}
\end{figure}

To determine the contribution of defect absorption to the internal loss rate of each fundamental cavity mode, we measure the transmission spectra of each mode at different input powers. 
The transmission spectra are normalized to the uncoupled fibre-taper transmission spectra (\Cref{fig:fig7_fibreEfficiency}\,(b)), and then fitted using \Cref{eq:transmissionDoubletTO}. 
The fits and data for each mode at three different input powers are plotted in \Cref{fig:fig9_SI_dataTransmissionKappas}, and each transmission spectrum is averaged using approximately five different scans. 
The modes manifest as either singlets or doublets, depending on the rate of back-scattering within the cavity ($g$) compared to the total loss rate ($\kappa$)~\cite{Borselli2005}.
Some of the modes also present Fano asymmetries~\cite{Suh2003} and power-dependent thermo-optic shifts\,\cite{Barclay2005,Itoi2025arXiv}; however, we are primarily interested in how the internal loss rate of each cavity mode changes with input power.
The internal loss rate, $\kappa_{\text{cav}}$, and external coupling rate of each mode at different powers are extracted from each fit and plotted in \Cref{fig:fig10_SI_dataKappas}\,(a) and (b), respectively. 
Note that the intrinsic linewidth is given by measurements of $\kappa_{\text{cav}}$ at low $\mathcal{N}_{\text{cav}}$, plotted in Fig.\ 3 and Fig.\ 10\,(a).

For most modes, $\kappa_{\text{cav}}$ is invariant with input power, with the exception of those at 979\,nm, 1047\,nm, and 1267\,nm.
In these modes, we observe that internal loss \textit{decreases} with the intracavity photon number, which evidences the presence of a two-level saturable absorber. 
This change in loss is not due to a change in the external coupling rate, $\kappa_{\text{ex}}$, as demonstrated by \Cref{fig:fig10_SI_dataKappas}\,(b)    . 
The external coupling rate of each mode remains constant with the number of intracavity photons. 
Instead, we attribute this to saturable absorption by a point defect.

Thermo-optic effects can distort the cavity resonance lineshapes. Such effects have been observed in microdisks with nanoscale diameter pedestals optimized for optomechanics\,\cite{lake2018optomechanically} and in photonic crystal nanocavities\,\cite{Itoi2025arXiv}. The microdisks used in our measurement, however, have relatively large pedestals with diameters $\sim 1\,\upmu\text{m}$, resulting in excellent thermal heat sinking. As a result, they exhibit little thermo-optic distortion to their lineshapes. More quantitatively, in our reported measurements the maximum change in temperature of devices, as inferred from these fits and the thermo-optic coefficient of diamond~\cite{Fontanella1977}, is approximately $2$~K (obtained at $P = 3\,\text{mW},\, \lambda = 1322\,$nm). The corresponding change in resonance wavelength is $2$~GHz, which is small compared to the linewidths measured here. 

\begin{figure}[t!]
    \centering
    \includegraphics{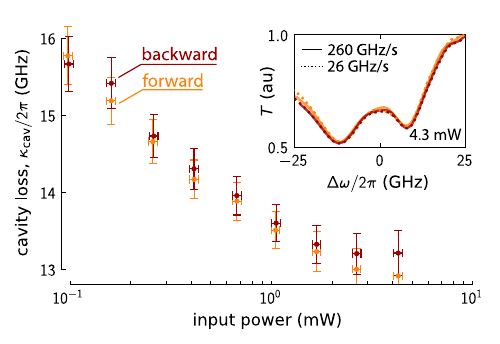}
    \caption{\textbf{Robustness to thermo-optic effects.}     
    Internal cavity loss $\kappa_{\text{cav}}$ is extracted as a function of power input to the fibre-taper from transmission spectra of the 1047\,nm mode. We use a scan rate of 260\,GHz/s in forward (small to large wavelength) and backward scan directions. 
    Additionally, the transmission profiles for the high power forward and backward scans at both 26\,GHz/s and 260\,GHz/s are shown in the inset.  
   }
    \label{fig:fig11_SI_scanSpeed}
\end{figure}

    To test whether thermo-optic effects are impacting the results, we perform two control measurements.
    First we vary scan direction, and second we vary scan speed. 
    The former allows us to determine if hysteresis effects related to thermo-optic nonlinearities affect the lineshapes. 
    The latter is used to verify whether we are operating in a quasi-static regime.
    We scan both forward and backward across the 1047~nm WGM at a speed of 260\,GHz/s and extract $\kappa_{\text{cav}}$ as a function of input power, as shown in \Cref{fig:fig11_SI_scanSpeed}.
    In both measurement directions, $\kappa_{\text{cav}}$ is seen to nonlinearly decrease with input power, and we do not observe any hysteresis effects. 
    Second, we measure the resonance at high power using two different scan speeds: 26\,GHz/s and 260\,GHz/s.
    The lineshapes of each scan, shown in the inset of \Cref{fig:fig11_SI_scanSpeed}, are in good agreement with one another.
    Thus, thermo-optic effects do not significantly impact the observed intensity dependent change in transmission resonance contrast and lineshape. 
    These measurements were all completed without use of optical amplification in order to demonstrate that amplifier noise has no effect on the observed relationship between $\kappa_{\text{cav}}$ and incident power.

    We also note that the weak thermo-optic effects present here have no impact on the transmission minimum measured on-resonance, which can be used to independently monitor $\kappa_{\text{cav}}$. For the under-coupled modes in our system, transmission on-resonance is governed by  $K = \kappa_{\text{ex}}/\kappa_{\text{cav}}(I)$ through $T_{\text{min}}(I) = (1-K)^2/(1+K)^2$\,\cite{Spillane2003,Barclay2005}. 
    The resulting decrease of $T_{\text{min}}(I)$ with increasing $I$ can be clearly seen from data in Figs.\,3 and 9. This provides a linewidth-independent confirmation that the cavity's internal loss decreases with increasing input power.
    Furthermore, note that this observed decrease in resonance contrast (smaller $\kappa_{\text{ex}}/\kappa_{\text{cav}}$) with increasing input power rules out optical forces changing the fibre position, which would be expected to increase $\kappa_{\text{ex}}$ and increase the resonance contrast by pulling the fibre taper closer to the microdisk.

\section{Saturable absorption}
\label{sec:DeriveSatAbs}
In a cavity, the energy loss rate due to linear absorption by defects in the material is proportional to the absorption coefficient, $\alpha$\,\cite{Saleh2019}:
\begin{align}
    \frac{\kappa_{\text{abs}}}{2\pi}= v_{\text{g}}\alpha \approx \frac{c}{n_{\text{g}}}\alpha\,,
    \label{eq:kappaAbstoAbsCoeffApp}
\end{align}
where $ v_{\text{g}}$ and $ n_{\text{g}}$ are respectively the group velocity and group index of the mode, and the approximation holds in regions of low dispersion.
The group index can be retrieved from simulations, and can be used to convert the measured temporal energy loss rates into a material-dependent absorption coefficient, $\alpha$. 
In turn, the absorption coefficient is proportional to the absorption cross-section at frequency $\omega$ of a single absorber, $\sigma_{\omega}$, the density of absorbers, $M$, and the intensity-dependent fractional ground state population density of the absorbers, $m_{0}$\,\cite{Boyd2008}:
\begin{align}
    \alpha =M  m_{0}\sigma_{\omega} = \alpha_0m_0\,.
    \label{eq:absCoefficientApp}
\end{align}

To determine the intensity-dependent absorption of the defect, we model it as a two-level-system, whose energy levels are separated by $E = \hbar\Omega$.
Off-resonant phonon-assisted absorption of photons with energy greater than $\hbar\Omega$ can incoherently excite the ground state population $m_0$ to the excited state population $m_1$, and is described by the following rate equations\,\cite{Rao:2022}:

\begin{align}
\frac{d m_0}{d t}&=-\frac{\sigma_{\omega} \langle I\rangle}{\hbar \omega}m_0+\frac{m_1}{\tau}\, 
\label{eq:ratem1}\\
\frac{d m_1}{d t}&=\frac{\sigma_{\omega} \langle I\rangle}{\hbar \omega}m_{0}-\frac{m_1}{\tau}\,,
\label{eq:ratem0}
\end{align}
where $\tau$ is the lifetime of the excited state, $\sigma_{\omega}$ is the absorption cross-section at frequency $\omega$, and $\langle I\rangle$ is the average intensity inside the diamond cavity at resonance.
Given population conservation ($m_0 + m_1 = 1$) in conjunction with \Cref{eq:ratem0,eq:ratem1}, the steady-state fractional population density of each state can be expressed as
\begin{align}
    m_0 &= \frac{1 }{1 + \frac{\langle I\rangle}{I_{\text{sat}}}} 
    \label{eq:m0}\\
    m_1 &= \frac{\frac{\langle I\rangle}{I_{\text{sat}}}}{1 + \frac{\langle I\rangle}{I_{\text{sat}}}}\,, \label{eq:m1}
\end{align}
where $I_{\text{sat}} = \hbar\omega/\sigma_{\omega}\tau$ is the saturation intensity of the two-level-system. 
We then use \Cref{eq:absCoefficientApp,eq:m0} to express the cavity absorption coefficient in terms of the absorption cross-section, absorber density, and saturation intensity:
\begin{align}
    \alpha = \frac{M\sigma_{\omega}}{1 + \frac{\langle I\rangle}{I_{\text{sat}}}}\,.
    \label{eq:absorptionCoeffApp}
\end{align}
Here, $\langle I\rangle$ is calculated using \Cref{eq:averageIntensity2} under the assumption that the laser is on resonance with the cavity mode ($\mathcal{N}_{\text{cav}} = \mathcal{N}_{\text{cav}}(\Delta\omega = 0)$). 
This treatment is an approximation, as it neglects both the dependence of the intracavity intensity—and therefore the absorption—on cavity detuning, as well as the spatial variation of the intensity within the cavity (see \Cref{sec:avgCavityIntensity}). 
Implicitly, \Cref{eq:absorptionCoeffApp} assumes that the intensity is roughly uniform over the absorbing material, which is approximately true in the case of the cavity studied here.
In the limits of strong under-saturation and strong over-saturation, this approximation is well justified because the absorption loss rate is effectively independent of intensity. 
Consequently, the simplifications used in the model do not significantly affect the extracted linear absorption coefficient, $\alpha_0$. 
However, they may introduce additional uncertainty in the extracted saturation intensity. 
This is not accounted for in the numerical uncertainty of the saturation intensity values presented in \Cref{tab:results}.

Combining \Cref{eq:kappaInt,eq:kappaAbstoAbsCoeffApp,eq:absCoefficientApp},we get 
\begin{align}
    \frac{\kappa_{\text{cav}}}{2\pi}= \frac{\kappa_{\text{i}}}{2\pi} + \frac{\kappa_{\text{p}}}{2\pi} + \frac{c}{n_g}\frac{M\sigma_{\omega}}{1 + \frac{\langle I\rangle}{I_{\text{sat}}}}\,.
    \label{eq:fitSatAbsApp}
\end{align}
To extract values for the saturation intensity and absorption parameters, this expression can be fit to $\kappa_{\text{cav}}(\langle I\rangle)$ extracted from experimental measurements, as shown in Fig.\ \ref{fig:fig4__Saturation}.

\section{Considerations of uncertainty}
\label{sec:uncertainty}
\begin{figure}[b!]
    \centering
    \includegraphics{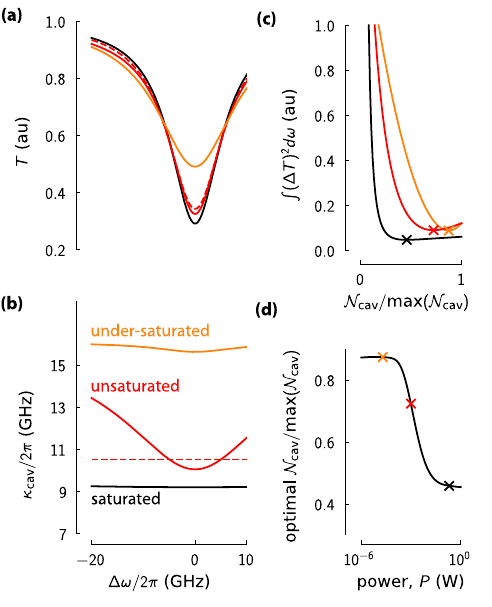}
    \caption{
    \textbf{Extracting the cavity photon number.}  
    \textbf{(a)} Here we plot the transmission spectra for three different input powers using a dynamic model which accounts for detuning-dependent loss (solid lines). 
    \textbf{(b)} Cavity loss for each of the three input powers is detuning-dependent and is most significant in the intermediate unsaturated regime between under-saturated and saturated absorption. 
    The corresponding prediction from the static model used to fit the experimental data is shown for comparison (dashed line).  
    \textbf{(c)} For a given input power, an experimentally extracted $\kappa_{\text{cav}}$ can be associated with an effective intracavity photon number by minimizing the difference between the dynamic and static transmission models.
    \textbf{(d)} The effective intracavity photon number corresponding to the fitted loss rate varies between approximately $0.45\max(\mathcal{N}_{\mathrm{cav}})$ and $0.85\max(\mathcal{N}_{\mathrm{cav}})$, depending on the saturation regime and the coupling strength. 
    This gives us an effective uncertainty range for $\mathcal{N}_{\mathrm{cav}}$ at a given input power.
    }
    \label{fig:fig12_extractionofphotonnumber}
\end{figure}

Determination of the saturation intensities and absorption coefficients is achieved by fitting several equations derived from the absorption and cavity transmission models introduced in the manuscript, such as \Cref{eq:fitSatAbs,eq:transmissionDoubletTO,eq:intracavityPhotonNumber}.
In this section, we detail these calculations and discuss how uncertainties in the measured and extracted parameters propagate into uncertainties in the inferred saturation intensities and absorption coefficients.

We first extract the cavity parameters, including the intrinsic and external loss rates, by fitting the measured transmission spectra (\Cref{fig:fig9_SI_dataTransmissionKappas}) using \Cref{eq:transmissionDoubletTO}. 
Parameter uncertainties are estimated from the covariance matrix obtained using the least-squares fitting procedure. 
These uncertainties include contributions from detector and amplifier noise, where applicable, and each transmission spectrum is averaged over five measurements to mitigate the effects of white noise from the detector, laser, and the optical amplifier.
The resulting uncertainties in the extracted loss rates are shown throughout the manuscript as error bars, including in \Cref{fig:fig4__Saturation,fig:fig10_SI_dataKappas}. 

An important limitation of the transmission spectrum fitting procedure is that the fitting model assumes a single detuning-independent cavity loss rate for a given input power. 
Physically, however, the loss is detuning-dependent because the intracavity intensity varies across the resonance and the absorber saturates nonlinearly with intensity. 
As a result, the cavity loss evolves continuously with laser detuning.

This effect is illustrated in \Cref{fig:fig12_extractionofphotonnumber}~(a,b), where we compare the cavity loss rates and corresponding transmission spectra generated using a dynamic model that explicitly accounts for detuning-dependent loss. 
Specifically, the dynamic model self-consistently solves for a detuning-dependent saturable loss rate that depends on the number of intracavity photons: $\kappa_{\text{cav}}(\Delta\omega) = \kappa_{\text{i}} + \kappa_{\text{p}} + \kappa_{\text{abs}}/(1 + \mathcal{N}_{\text{cav}}/\mathcal{N}_{\text{sat}})$. 
The arbitrary choice of $N_{\text{sat}}$ for the illustration in \Cref{fig:fig12_extractionofphotonnumber} does not impact the resulting uncertainties.
At high input power, the absorbers are strongly saturated over the entire resonance and the loss is low and approximately constant with detuning.  
Similarly, at low power the absorber remains largely under-saturated, again producing high, nearly detuning-independent loss.  
The largest discrepancy occurs in the intermediate saturation regime. Near resonance, the intracavity intensity is high, and the loss is partially saturated. Far from resonance, however, the intracavity intensity is low, and the loss is under-saturated.
Consequently, the measured transmission spectrum cannot be rigorously described by a single static loss rate.
Despite this limitation, the static model still provides an effective loss parameter that captures the overall resonance lineshape and enables robust fitting of the experimental data. 
We therefore numerically compare the dynamic and static models to determine the effective intracavity photon number associated with the effective cavity loss rate.

For a given input power, we first use the dynamic model to calculate the transmission spectrum $T_{\text{d}}(\Delta\omega)$ and the corresponding $\kappa_{\text{cav}}(\Delta\omega)$.
We then generate a family of static-model transmission spectra $T_{\text{s}}(\Delta\omega, \kappa_{\text{cav}})$ using different constant values of $\kappa_{\text{cav}}$.
For each static spectrum, we compute the integrated squared difference
\begin{align}
\int(\Delta T)^2 d(\Delta\omega) = \int \left[T_{\mathrm{d}}(\Delta\omega)-T_{\mathrm{s}}(\Delta\omega)\right]^2 d(\Delta\omega)\,,
\end{align}
as illustrated in \Cref{fig:fig12_extractionofphotonnumber}~(c). 
Minimizing this quantity identifies the static loss rate that best reproduces the dynamic transmission spectrum. 
This procedure allows us to map the experimentally extracted $\kappa_{\text{cav}}$ to an effective $\mathcal{N}_{\text{cav}}$, and it allows us to more accurately assess the uncertainty in  $\mathcal{N}_{\text{cav}}$.
Using this approach over a range of input powers we find that the effective $ \mathcal{N}_{\text{cav}}$ varies between approximately $0.45\max(\mathcal{N}_{\text{cav}})$ and  $0.85\max(\mathcal{N}_{\text{cav}})$. 
This spread constitutes the dominant source of uncertainty in the extracted saturation intensity, which we account for using \Cref{eq:intracavityPhotonNumber} by estimating $\mathcal{N}_{\text{cav}}$ as:
\begin{align}
    \mathcal{N}_{\text{cav}}(\kappa_{\text{cav}}) &= 0.65\operatorname{max}\left[\frac{P\sqrt{\eta_{\text{fibre}}}}{\hbar \omega}
    \left(\frac{\langle\hat{a}(\Delta\omega)\rangle}{\langle\hat{s}\rangle}\right)^2\right]\,,
\end{align}
and its uncertainty as
\begin{align}
\begin{aligned}
    \delta \mathcal{N}_{\text{cav}} 
    & = \frac{1}{\hbar \omega}\left(\frac{\langle\hat{a}(\Delta\omega)\rangle}{\langle\hat{s}\rangle}\right)^2 \times \\ 
    & \left( \frac{P^2\delta \eta_{\text{fibre}}^2}{4\eta_{\text{fibre}}} + \eta_{\text{fibre}}\delta P^2\right)^{\frac{1}{2}} + 0.2\mathcal{N}_{\text{cav}}\,.
\end{aligned}
\end{align}
Here, the first term accounts for standard propagation of parameter uncertainty, and the final term accounts for the uncertainty introduced by associating the fitted static loss rate with an effective intracavity photon number. 
These calculations ignore contributions from uncertainties in the extracted loss rates, which are small compared to those from uncertainty in $\mathcal{N}_{\text{cav}}$, the transmission efficiency $\eta_{\text{fibre}}$, and the input optical power $P$.
Note that although $\delta\mathcal{N}_{\text{cav}}$ ignores contributions from loss rate uncertainties, the orthogonal distance regression procedure used to extract the absorption coefficient and the saturation intensity does include contributions from uncertainty in loss rate.

The above approach allows us to extract the cavity loss rates, the corresponding intracavity photon numbers, and each of their uncertainties.
Then, using \Cref{eq:averageIntensity2}, we calculate the effective average intensity, which also depends on the group index and the effective mode volume. 
In our case, however, $\delta \mathcal{N}_{\text{cav}}$ is much larger than the uncertainty in the simulation-extracted group index and effective mode volume.
Therefore, we only propagate uncertainty from the intracavity photon number. 
Similarly, the absorption coefficient, extracted by fitting \Cref{eq:fitSatAbs}, is dominated by uncertainty in the extracted loss rates, and we therefore ignore contributions from uncertainty in the group index.
To account for uncertainties in the fit to \Cref{eq:fitSatAbs}, we use an orthogonal distance regression technique which accounts for uncertainty in $\langle I \rangle$ and in $\kappa_{\text{cav}}$. 
The uncertainty in the extracted parameters ($\alpha_0$ and $I_{\text{sat}}$) is visualized by the broad translucent lines shown in \Cref{fig:fig4__Saturation}.



\bibliography{bibliographies/main}

\end{document}